\documentclass[12pt]{iopart}

\usepackage{graphicx}
\usepackage{subfigure}  
\begin{document}

\title[]
{Electromagnetically induced transparency in \\ cold $^{85}$Rb atoms trapped in the ground hyperfine \\ F = 2 state }

\author{V. B. Tiwari, S. Singh, H. S. Rawat, Manoranjan P. Singh,\\ S. C. Mehendale}

\address{Laser Physics Applications Division, \\
Raja Ramanna Centre for Advanced Technology,\\
Indore 452013, India}
\ead{vbtiwari@rrcat.gov.in}
\begin{abstract}

We report electromagnetically induced transparency (EIT) in cold $^{85}$Rb atoms, trapped in the lower hyperfine level F = 2, of the ground state 5$^{2}S_{1/2}$ (Tiwari V B \textit{et al} 2008 {\it Phys. Rev.} A {\bf 78} 063421). Two steady state $\Lambda$-type systems of hyperfine energy levels are investigated using probe transitions into the levels F$^{\prime}$ = 2 and F$^{\prime}$ = 3 of the excited state  5$^{2}P_{3/2}$ in the presence of coupling transitions F = 3 $\rightarrow$ F$^{\prime}$ = 2 and  F = 3 $\rightarrow$ F$^{\prime}$ = 3, respectively. The effects of uncoupled magnetic sublevel transitions and coupling field's Rabi frequency on the EIT signal from these systems are studied using a simple theoretical model. 

\end{abstract}


\section{ Introduction }
\label{Introduction}
Electromagnetically induced transparency (EIT) in atomic systems is a quantum interference effect resulting in reduced absorption of a weak probe field in resonance with an atomic transition while propagating through a medium in presence of a strong coupling field \cite{Harris1997}. Since its first experimental realization in 1991 by Boller et al. \cite{Boller1991} in strontium vapour, the phenomenon of EIT has been instrumental in the progress of various fields, such as nonlinear optics \cite{Yan2001-1, Kang2003, Braje2004}, sensitive magnetometry \cite{Lee1998}, Rydberg states spectroscopy \cite{Mauger2007,Weatherill2008} and laser frequency stabilization \cite{Moon2004,Ray2007}. Mostly, these EIT related studies have been carried out either using Doppler-free configuration of coupling and probe beams in an atomic vapour cell \cite{Li1995,Moon1999,Wei2005} or in an inherently Doppler-free environment of cold atoms \cite{Hopkins1997,Cataliotti1997,Durrant1998}. The elimination of Doppler broadening effect and low collisional dephasing rate has made the cold atoms an attractive medium to explore the EIT effect with moderate coupling beam intensities \cite{Clarke2001}. The use of cold atoms, in addition to simplifying the analysis of the results, has also allowed a much more flexible experimental arrangement due to the possibility of propagation of coupling and the probe beams in arbitrary directions. As a result, many recent experiments on quantum information processing \cite{Lukin2001,Liu2001} and quantum interference switching \cite{Braje2003,Kang2006,Zhang2007} have successfully involved cold alkali atoms using either three-level atomic configurations such as $\Lambda$-type, $V$-type, ladder-type or their combinations \cite{Fleischhauer2005}. The strength of the signal in an EIT experiment usually depends on the atomic configuration used. For example, a three-level $\Lambda$-type system with minimum coherence dephasing rate of the dipole-forbidden transition provides maximum transparency at the line centre for the EIT signal \cite{Fulton1995}. A practical $\Lambda$-type system such as that realized with alkali atoms, however, involves hyperfine energy levels with many associated degenerate Zeeman-sublevels. Chen et al. \cite{Chen2000} have observed that the spectral profile of the EIT signal depends on the individual transitions among the Zeeman-sublevels which can be grouped together according to the selection rules to form a multi-fold of $\Lambda$-type systems. Yan et al. \cite{Yan2001-2} have further noticed that an ideal closed three-level $\Lambda$-type system can be obtained from such a laser-coupling formation, provided, the Zeeman-sublevels of the excited hyperfine level F$'$ are simultaneously coupled to the Zeeman-sublevels of ground coupling hyperfine level F$_{\rm{c}}$ and ground probe hyperfine level F$_{\rm{p}}$. Using a dark-MOT configuration \cite{Ketterle1993}, this formation was realized using cold $^{87}$Rb atoms collected in the lower hyperfine level of the ground state and satisfying the condition, F$_{\rm{c}}$ $>$ F$_{\rm{p}}$. However, the strong coupling and trapping laser beams operating at D$_{1}$-line and D$_{2}$-line transitions, respectively, involved the same upper hyperfine level of the ground state. This required the experiment to be operated in a sequential mode with a periodic turning-off of the trapping beams.
\\In this work, we present a new experimental scheme to obtain $\Lambda$-type EIT systems using cold $^{85}$Rb atoms in the lower hyperfine level F = 2 of the ground state 5$^{2}S_{1/2}$. Here, the cold atoms were obtained by directly cooling and trapping them into the lower hyperfine ground state without resorting to the dark-MOT technique. This magneto-optical trapping of $^{85}$Rb atoms involving cycling transition F = 2 $\to$ F$^{\prime}$ = 1 has been reported earlier by us \cite{Tiwari2008}. In contrast with the transient mode of operation reported in Ref. \cite{Yan2001-2}, our scheme allowed us to use a steady state cold atomic sample in the lower hyperfine level of the ground state. Moreover, the formation of $\Lambda$-type EIT systems did not involve D$_{1}$-line coupling and probe excitations resulting in a considerable simplification of the experimental setup. Using mutually perpendicular linearly polarized coupling and probe beams, we obtained two steady state $\Lambda$-type EIT systems when cold atoms were probed for transition into the hyperfine level F$^{\prime}$ = 2 and F$^{\prime}$ = 3 of the excited state  5$^{2}P_{3/2}$ in the presence of coupling transitions F = 3 $\rightarrow$ F$^{\prime}$ = 2 and F = 3 $\rightarrow$ F$^{\prime}$ = 3, respectively. The experimental observations on the effects of uncoupled magnetic sublevel transitions and coupling field's Rabi frequency on the EIT signal from these systems were found to be in good agreement with the theoretical calculations. These results clearly demonstrate the role of Zeeman sublevels associated with the common excited state in the consideration of a $\Lambda$-type EIT system involving hyperfine levels. 


\section{ Experimental }
\label{exp}
Figure 1(a) shows the schematic diagram of our experimental setup. The MOT chamber was pumped down to a pressure of $\sim $ 1$\times $10$^{{\rm -} {\rm 8}}$ torr using a turbo molecular pump and a sputter-ion pump. Rb vapour was injected in the chamber by passing a constant current of ~$\sim $ 3.5 A through two Rb-getters fixed in series. The magnetic field gradient required at the center of the trap was produced by a pair of quadrupole coils. During the experiment, the radial magnetic field gradient dB/dz was fixed at 10 G/cm. All stray magnetic fields were removed by the use of three orthogonal pairs of compensating coils. Three orthogonal pairs of $\sigma^{+}$ and $\sigma^{-}$ trapping laser beams (T) with nearly equal intensity were produced using PBS with $\lambda $/2 and $\lambda $/4 retardation plates. One pair of counterpropagating laser beams was aligned with the axis of the coils (x-axis). The other two pairs of counterpropagating beams were located in the symmetry plane (yz-plane) of the coils, inclined at $45^{0}$ with respect to the vertical z-axis.  These trapping laser beams of nearly Gaussian shape with a 1/e$^{{\rm 2}}$ diameter of 7.6 mm, intersected near the zero of the magnetic field. The frequency of the trapping laser was locked at the centre of the cyclic transition 5$^{{\rm 2}}$S$_{{\rm 1}{\rm /} {\rm 2}{\rm }{\rm} }$(F = 2)$ \to $5$^{{\rm 2}}$P$_{{\rm 3}{\rm /} {\rm 2}{\rm} }$ (F$^{\prime}$= 1) of $^{{\rm 8}{\rm 5}}$Rb using the bi-polarization spectroscopy technique \cite{Tiwari2006} at $\lambda$ =780 nm. The total intensity of all the six trapping laser beams used in our experiment was 50 mW/cm$^{{\rm 2}}$( the saturation intensity I$_{{\rm s}}$  $\sim $ 2.8 mW/cm$^{{\rm 2}}$ for the cooling transition). A linearly polarised repumping laser beam (R) separately propagated through the trap center along the vertical z-axis with its frequency locked to the peak of 5$^{{\rm 2}}$S$_{{\rm 1}{\rm /} {\rm 2}{\rm} {\rm} }$(F = 3)$ \to $5$^{{\rm 2}}$P$_{{\rm 3}{\rm /} {\rm 2}{\rm} }$ (F$^{\prime}$= 3 or 2) transition using the Doppler-free dichroic lock technique \cite{Wasik2002}. The repumping laser beam had a 1/e$^{{\rm 2}}$ diameter of 5.6 mm with maximum power of 20 mW. In these experiments, the repumping beam of the trap was also used as a coupling beam C. A weak probe beam P of power 1 $\mu$W with nearly uniform intensity was passed through the atom cloud and its transmission was detected using photo-detector PD-1. The two beams C and P have orthogonal linear polarizations and propagate perpendicular to each other. The mutual perpendicular propagation of these beams ensured that any observed EIT signal was produced by the cold atoms and not by the thermal background rubidium vapour. The fluorescence of the cloud was measured using a low noise level photo-detector PD-2 and a CCD camera to estimate the number and temperature of the trapped atoms. The energy level diagram used for investigating three-level $\Lambda$-type system with $^{85}$Rb atoms trapped in the lower hyperfine ground state is shown in Fig.1 (b). The frequency of the trapping laser was red detuned by $\Delta _{{\rm L}}/2\pi$ = -12 MHz, from the centre of the cyclic transition 5$^{2}$S$_{1/2}$(F = 2)$ \to $5$^{{\rm 2}}$P$_{{\rm 3}{\rm /}{\rm 2}{\rm} }$ (F'= 1) of $^{{\rm 8}{\rm 5}}$Rb by using a pair of accousto-optical modulators. In the $\Lambda$-type system studied here, the hyperfine level F$^{\prime}$= 3 (or F$^{\prime}$= 2) of 5$^{{\rm 2}}$P$_{{\rm 3}/{\rm 2}}$ served as excited state $\left|2\right\rangle$. The hyperfine levels F = 2 and F = 3 of the ground state 5$^{\rm 2}$S$_{{\rm 1}/{\rm 2}}$ were used as two ground states $\left|1\right\rangle$ and $\left|3\right\rangle$ of the $\Lambda$-type system, respectively [Fig. 2 (a) and (b)]. The atoms were pumped between the 5$^{\rm 2}$S$_{{\rm 1}/{\rm 2}}$, F = 3 and 5$^{{\rm 2}}$P$_{{\rm 3}/{\rm 2}}$, F$^{\prime}$= 3 (or F$^{\prime}$= 2) states by the coupling laser (solid lines) and probed between the 5$^{\rm 2}$S$_{{\rm 1}/{\rm 2}}$, F = 2 and 5$^{{\rm 2}}$P$_{{\rm 3}/{\rm 2}}$, F$^{\prime}$= 3 (or F$^{\prime}$= 2) states by the weak probe laser (dotted lines). 


\section{ Theoretical Analysis}
\label{th}   
\indent To understand the experimentally obtained EIT signal, we have used a theoretical model for three-level $\Lambda$-type system described by a semiclassical theory \cite{Li1995,Scully1997}. In this three level model, the energy level $\left|1\right\rangle$ is probed for transition to an upper energy level $\left|2\right\rangle$ which itself is strongly coupled to another lower level $\left|3\right\rangle$. The transitions $\left|1\right\rangle$ $\rightarrow$ $\left|2\right\rangle$  and $\left|3\right\rangle$ $\rightarrow$ $\left|2\right\rangle$ have the resonance frequencies as $\omega_{21}$ and $\omega_{23}$, respectively. A weak probe laser beam having frequency $\omega_{p}$ scans over the excited state hyperfine level which is also coupled to the another hyperfine level of the ground state by a resonant strong coupling laser beam. 
\\For a three level $\Lambda$-type system composed of hyperfine levels, the degenerate Zeeman sublevels play an important role in deciding the strength of the EIT signal \cite{Chen2000}. As long as all the Zeeman probe transitions are connected with the coupling transitions, the $\Lambda$-type configuration can be treated as an effectively closed system. This system may be taken as the superposition of several independent three level systems \cite{Wei2005}. The resulting EIT signal is also maximum for this case. On the other hand, presence of some probe transitions without the interference of the coupling field renders the EIT system, effectively not closed. Consequently such a system can be modeled by a combination of three level and two level systems. The resulting spectrum is a combination of the EIT profile and the Lorentzian absorption profile with a peak at the resonance frequency. Since the spatial variation of magnetic field over the atom cloud dimensions was small ($\sim$ 1G), we have assumed the hyperfine levels to be nearly degenerate. In this approximation, the decoherence rate due to the Zeeman splitting of the hyperfine levels can be added to the decoherence rate of the relaxation process containing spontaneous decay \cite{Chen2000}. Under this approximation, the real and imaginary parts of the complex succetibility, $\chi$ = ($\chi'$ + $i\chi''$) for a resonant coupling field are given by,
\begin{eqnarray}
\chi' = \Delta_{p}\cdot\frac{n_{a}}{\epsilon_{0} \hbar }\cdot\left[\frac{x-\gamma_{31}\:y}{x^{2}+\Delta_{p}^{2}\:y^{2}}\sum\limits_{i,j}d_{ij}^{2} + \frac{1}{\gamma_{21}^{2}+\Delta_{p}^{2}} \sum\limits_{k,l}d_{kl}^{2}\right]
\end{eqnarray}
\begin{eqnarray}
\chi'' = \frac{n_{a}}{\epsilon_{0} \hbar }\cdot\left[\frac{\gamma_{31}x + \Delta_{p}^{2}\:y}{x^{2}+\Delta_{p}^{2}\:y^{2}} \sum\limits_{i,j}d_{ij}^{2}+ \frac{\gamma_{21}}{\gamma_{21}^{2}+\Delta_{p}^{2}} \sum\limits_{k,l}d_{kl}^{2}\right]
\end{eqnarray}
where, $\Delta_{p}$ = $\omega_{p}$ - $\omega_{21}$, x = $(\Omega_{c}^2/4 - \Delta_{p}^{2}) + \gamma_{21}\gamma_{31}$ and y = $(\gamma_{21} + \gamma_{31})$. 
\\Here, n$_{a}$ is the number density of cold atoms and $\rm {d_{ij}}$ and $\rm {d_{kl}}$ are the density matrix elements for the probe transition. The (i,j) pair refers to the probe transitions from magnetic sublevels of state $\left|1\right\rangle$ to the magnetic sublevels of excited state $\left|2\right\rangle$ which are also connected by the coupling beam. The summation over (k,l) pair refers to the probe transitions from magnetic sublevels of state $\left|1\right\rangle$ to the magnetic sublevels of excited state $\left|2\right\rangle$ which are not connected by the coupling beam. Accordingly, the first term inside the bracket on the right hand side of Eqs (1) and (2) denotes the EIT process whereas the second term in the respective equations corresponds to simple refraction and absorption of the probe beam. The Rabi frequency of the coupling light is given by, $ \Omega_{c} = \Gamma \times \sqrt{\frac{I_{c}}{2I_{s}}}$, where $I_{c}$  is the coupling laser beam intensity, $I_{s}$ is the saturation intensity and $ \Gamma $ (2 $\pi\times $ 5.9 MHz) is the natural line-width of the transition. 
\\The mean radiative decay rates of levels $\left|2\right\rangle$ and $\left|1\right\rangle$ is $\gamma_{21}$, whereas the non-radiative decay rate between the ground states, $\left|1\right\rangle$ and  $\left|3\right\rangle$ is $\gamma_{31}$. The strength of the EIT signal strongly depends on the coherence dephasing rate of the dipole-forbidden transition between energy states $\left|1\right\rangle$ and  $\left|3\right\rangle$. For an ideal EIT $\Lambda$-type system, $\gamma_{21}$ = $ \Gamma/2 $ and $\gamma_{31}$ = 0, results in total transparency at the transition line center. However, when the laser line shape is considered to be a Lorentzian shape, the effect of the finite laser line-width should be included in the decay rates. Moreover, the magnetic field inhomogeneity further adds to the decoherence rates for the relaxation processes. This magnetic field inhomogeneity depends on the magnetic field shifting of the magnetic sublevels over the cold atom cloud dimensions.  So that $\gamma_{21}\rightarrow \gamma_{21} + \gamma_{p}+ \gamma_{m}$ and $\gamma_{31}\rightarrow \gamma_{31} + \gamma_{p} + \gamma_{c} + \gamma_{m}$, where $\gamma_{p}$ and $\gamma_{c}$ are the line-widths of the probe and coupling beams, respectively and $\gamma_{m}$ is the Zeeman shift caused by the magnetic field in the region of atom cloud. 
\\The real and imaginary components of complex dielectric function are related with the components of the complex succeptibility, as follows, 
\begin{eqnarray}
\nonumber
K_{r} = 1 + \chi';  \  K_{i} = \chi''
\end{eqnarray}
The propagation of the probe electric field along the z-direction in a medium having complex refractive index n = n$_{r}$+$i$ n$_{i}$, is given by \cite{Reitz1979},
\begin{eqnarray}
\nonumber
E_{p}(r,t) = E_{0}e^{-k_{i}z} e^{-i(\omega t - k_{r}z)}
\end{eqnarray}
where the real and imaginary components of the complex propagation wavevector are given by k$_{r}$ = n$_{r}$ $\omega$/c and k$_{i}$ = n$_{i}$ $\omega$/c, respectively and therefore, the wave propagates with phase velocity c/$n_{r}$ and attenuation constant $n_{i}$$\omega$/c.
\\Using real and imaginary parts of the complex dielectric function, the components of complex refractive index can be written as ,
\begin{eqnarray}
\nonumber
n_{r}  = \sqrt{\frac{1}{2}\left(\textit{K}_{r} + \sqrt{\textit{K}^2_{r} + \textit{K}^2_{i}}\right)}
\end{eqnarray}
\begin{eqnarray}
\nonumber
n_{i} = \sqrt{\frac{1}{2}\left(-\textit{K}_{r} + \sqrt{\textit{K}^2_{r} + \textit{K}^2_{i}}\right)}
\end{eqnarray}
Therefore, the transmitted intensity of a probe beam having frequency $\omega$ = $\omega_{p}$ and passing through a cold atomic sample of length $l$ can be obtained from the relation,  
\begin{eqnarray}
I_{p} = I_{0}\; e^{- \alpha l}  
\end{eqnarray}
where $I_{0}$ is the incident intensity of the probe beam, and $\alpha$=2$n_{i}$$\omega_{p}$/c
is the intensity absorption coefficient. 
We have defined the normalized EIT peak height (EIT$_{\rm{peak}}$), as the ratio of absorption change of the probe beam in presence of coupling beam to the maximum absorption in the absence of the coupling beam.
\begin{eqnarray}
\rm{EIT}_{\rm{peak}} = \frac{(e^{- \alpha_{c} l} - e^{- \alpha_{0} l})}{(1 - e^{- \alpha_{0} l})}
\end{eqnarray}
where $\alpha_{c}$ and $\alpha_{0}$ are the absorption coefficients in presence and absence of the coupling beam, respectively. The results of calculation of EIT$_{\rm{peak}}$ and their comparison with experimental results are discussed in the next section. 

\section{ Results and discussion }
\label{result}
As mentioned earlier, we have investigated EIT signal from $\Lambda$-type systems obtained by tuning coupling laser frequency to the peak of two different transitions, 5$^{2}$S$_{1/2}$(F = 3) $\to$ 5$^{2}$P$_{3/2}$ (F$^{\prime}$= 3) and (F$^{\prime}$= 2), and shown in Fig.2(a) and (b), respectively. These $\Lambda$-type systems involve transitions among the Zeeman sublevels corresponding to ground and excited hyperfine levels. The strong coupling laser connects the state $\left|3\right\rangle$ and $\left|2\right\rangle$ whereas the weak probe laser was used to investigate the state $\left|1\right\rangle$. For orthogonal linearly polarized coupling and probe laser beams, it is easy to see whether each probed transition from a sublevel of state $\left|1\right\rangle$ to state $\left|2\right\rangle$ is coupled to a transition from a sublevel of state $\left|3\right\rangle$ to state $\left|2\right\rangle$.
A careful observation of the $\Lambda$-type system shown in Fig. 2(a), clearly suggests that it is not effectively closed. Here, the coupling transition, $F=3, m_{F}=0\rightarrow F'=3, m_{F'}=0$ is dipole forbidden. The probe transitions $F=2, m_{F}=-1\rightarrow F'=3, m_{F'}=0$ and $F=2, m_{F}=1\rightarrow F'=3, m_{F'}=0$ take place without the interference of the coupling field. In contrast, the $\Lambda$-type system shown in Fig.2(b) presents a closed $\Lambda$-type system. This configuration has all the sublevels of probed transition coupled to ground state sublevels by the strong coupling laser.  
\\Fig.3(a) shows the variation of the probe transmission, as the probe frequency was scanned from 5$^{2}$S$_{1/2}$(F = 2) across the 5$^{ 2}$P$_{3/2}$ hyperfine levels and the coupling beam was kept at exact resonance with the 5$^{2}$S$_{1/2}$(F = 3) $\to$ 5$^{2}$P$_{3/2}$ (F$^{\prime}$= 3 or F$^{\prime}$= 2) transitions in $^{85}$Rb. The transmitted probe intensity shows a narrow EIT peak at the centre of the absorption dip corresponding to the coupled level. The probe absorption measurements confirmed that more than 90 $\%$ of the $^{85}$Rb ground state atoms were in the F = 2 hyperfine level. The number density of trapped atoms was estimated to be about 2 $\times$ $10^{8}$ cm$^{-3}$ when the repumping beam frequency was resonanant with the F'= 3 level. We observed that the number density of the trapped atoms decreased to approximately by a factor of 4 when the resonance position of the repumping laser beam was shifted from the F = 3 $\to$ F$^{\prime}$= 3 transition to the F = 3 $\to$ F$^{\prime}$= 2 transition while keeping its intensity unchanged. This is in accordance with the different transition strengths associated with these excitations \cite{Metcalf1999}. The values of saturation intensity, I$_{{\rm s}}$, for a linearly polarized light field corresponding to F = 3 $\to$ F$^{\prime}$= 3 and F = 3 $\to$ F$^{\prime}$= 2 transitions, were estimated to be 4 mW/cm$^{{\rm 2}}$ and 17.4 mW/cm$^{{\rm 2}}$, respectively \cite{Steck2008}. The temperature of the atom cloud was estimated to be around 4.4 mK by using expansion images of the cloud. The full width at half maximum (FWHM) of the cold atom cloud  along the propogation directions of probe and coupling beams were measured to be 1.56 mm and 2.12 mm, respectively.    
Fig.3(b) presents a closer look of the observed EIT signals alongwith simulated ones from the theoretical model of a three level $\Lambda$-type system. The experimentally observed EIT signal agrees well with the theoretically calculated signals. For our theoretical calculations of both the complete and incomplete $\Lambda$-type systems, we have used $\gamma_{p}$ = $\gamma_{c}$ = 0.5 MHz. However, the value of decoherence due to the magnetic field inhomogeneity, $\gamma_{m}$, was estimated to be different for the two cases. The maximum Zeeman shift experienced by the excited and ground state Zeeman sublevels in the magnetic field over the atom cloud dimensions ($\sim$1G) corresponded to the value of $\gamma_{m}$ (see Fig.1(b)). For the incomplete $\Lambda$-type system as shown in Fig.2(a), $\gamma_{m}$ was estimated to be 2.6 MHz whereas for the complete $\Lambda$-type system as shown in Fig.2(b), it was estimated to be 0.4 MHz. The full width at minimum transmission of the EIT peak measures the average Rabi frequency experienced by the atoms \cite{Clarke2001}. Fig.4(a) demonstrates the linear dependence of experimentally observed average Rabi frequencies for coupling to F$^{\prime}$ = 3 and F$^{\prime}$ = 2 levels, on the square root of the coupling laser beam power. The power of the coupling laser beam was varied using calibrated neutral density filters. Fig.4(b) shows the variation of the normalized EIT peak height with the coupling Rabi frequency $\Omega_{\rm{c}}$, for the two $\Lambda$-type systems investigated. For the $\Lambda$-type system obtained from coupling to F$^{\prime}$ = 3 level, which is not closed,  EIT$_{\rm{peak}}$ was estimated to be $\sim$ 0.45 for $\Omega_{\rm{c}}$$\sim$ 10 MHz. Whereas, the EIT$_{\rm{peak}}$ was $\sim$ 0.9 for $\Omega_{\rm{c}}$$\sim$ 10 MHz, when an effectively closed $\Lambda$-type system was obtained by coupling to F$^{\prime}$ = 2 level. The normalized EIT peak height increased with the coupling Rabi frequency $\Omega_{\rm{c}}$ for both the $\Lambda$-type systems, as expected. However, the value of EIT$_{\rm{peak}}$ was observed to be comparatively much larger for the closed $\Lambda$-type system. These experimental results in good agreement with the theoretical calculations underline the importance of considering the Zeeman sublevels in the determination of the EIT signal from a $\Lambda$-type system involving hyperfine energy levels.

\section{ Conclusion }
\label{conclusion}
In summary, we have employed a new experimental scheme to obtain $\Lambda$-type EIT systems in a steady state sample of cold $^{85}$Rb atoms in the hyperfine level F = 2 of 5 $^{2} S_{1/2}$ ground state. The effects of uncoupled magnetic sublevel transitions and coupling field's Rabi frequency on the EIT signal from these systems have been studied. The  results clearly show that the EIT signal from a $\Lambda$-type system involving hyperfine levels depends on the Zeeman sublevels of the common excited state. We expect that our alternative experimental scheme to obtain a steady-state $\Lambda$-type EIT system with cold atoms in the lower hyperfine level may find interest in the experiments related to light pulse storage \cite{Peters2009} and optical switching devices with cold atoms \cite{Bajcsy2009,Dawes2009}.

\newpage

\begin{figure}[htb]
\begin{center}
\subfigure[]{
\includegraphics[height=6cm,width=7cm]{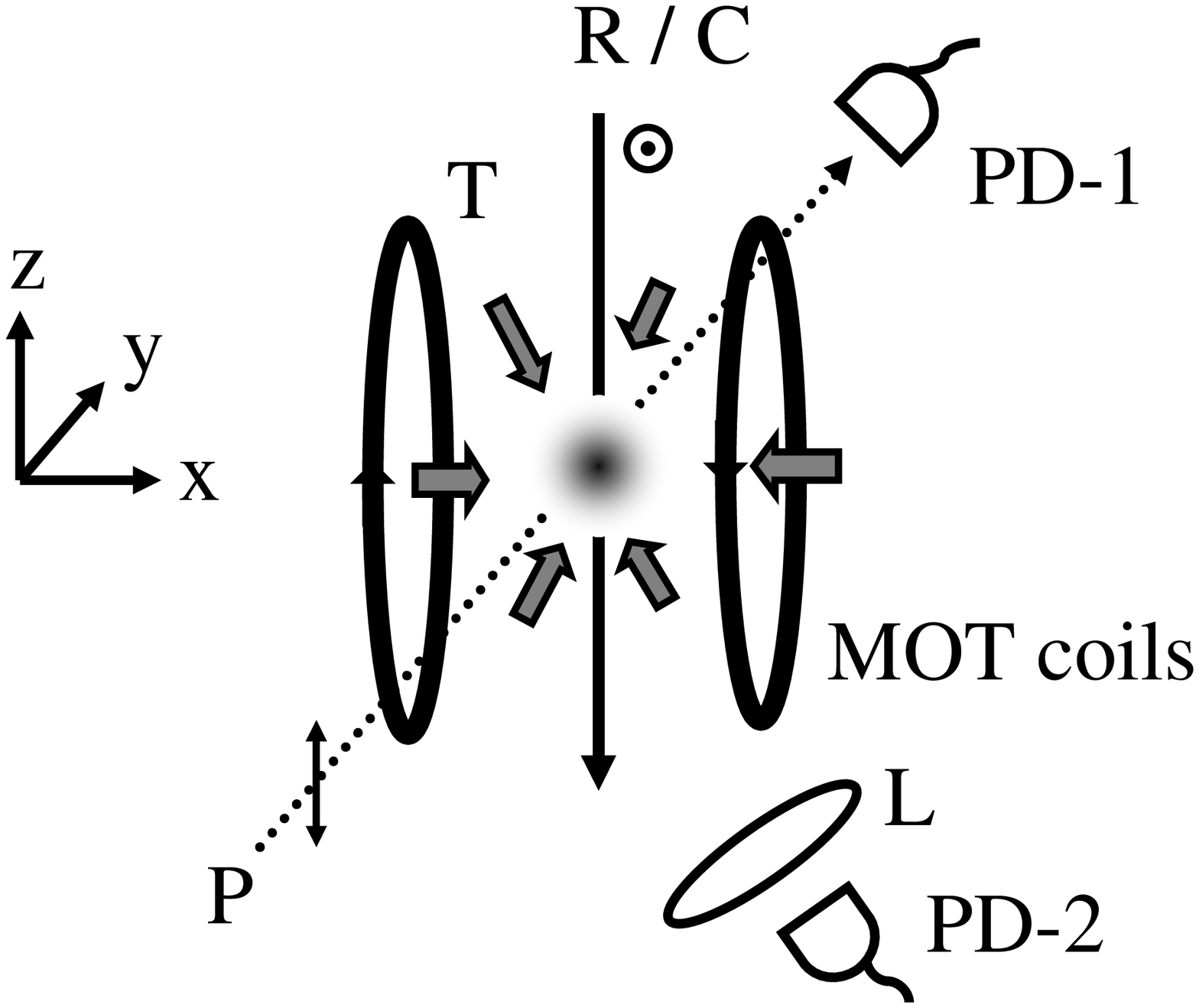}}
\hspace{0.5cm}
\subfigure[]{
\includegraphics[height=6cm,width=7cm]{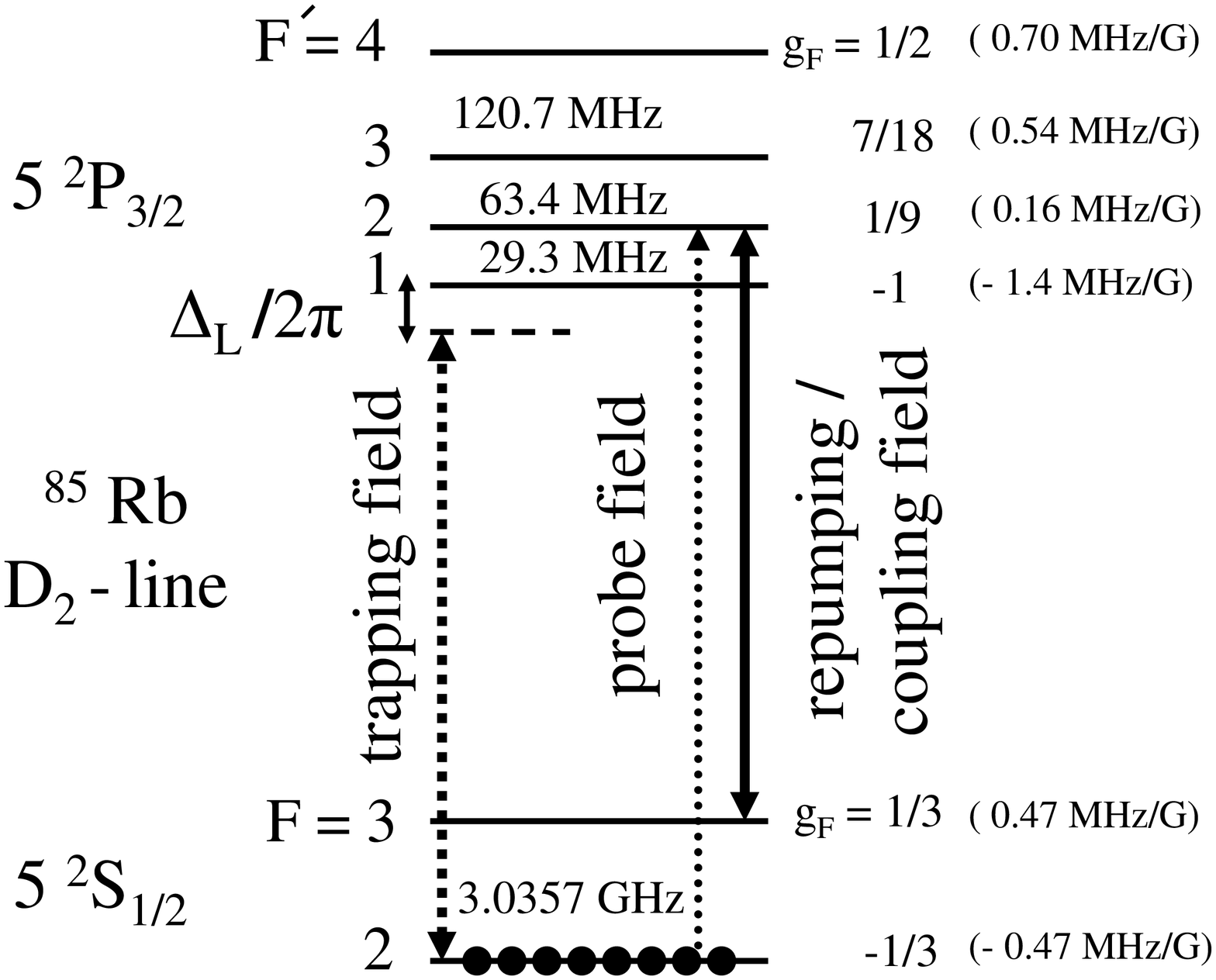}}
\end{center}
\caption{(a) Schematic diagram of the experimental setup. T: MOT trapping laser beams; R/C: repumping laser beam acting as a coupling beam; PD-1 and PD-2: photodiodes; L: collecting lense. (b) $\Lambda$-type system with cold atoms trapped in F = 2 state. The approximate Land$\acute{\rm{e}}$ $\rm{g_{F}}$-factors for each hyperfine level are given with the corresponding Zeeman splittings between adjacent magnetic sublevels. }
\end{figure}

\newpage

\begin{figure}[htb]
\begin{center}
\includegraphics[height=7cm,width=9cm]{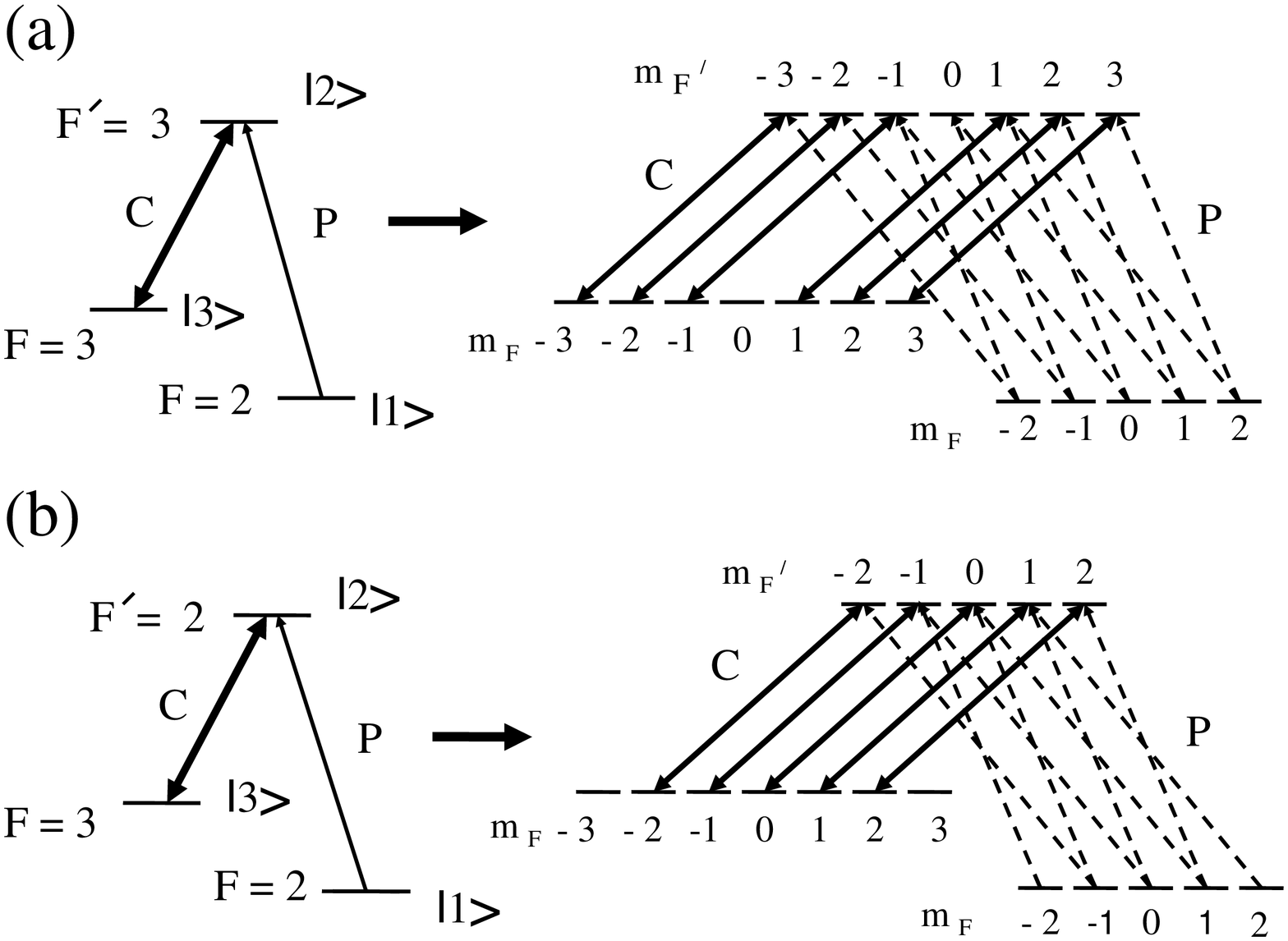}
\end{center}
\caption{(a) Incomplete $\Lambda$-type system (b) complete $\Lambda$-type system with cold atoms trapped in F = 2 state alongwith the  detailed coupling among the Zeeman-sublevels with orthogonal linear polarizations of coupling and probe fields.}
\end{figure}

\newpage

\begin{figure}[htb]
\begin{center}
\subfigure[]{
\includegraphics[height=6.5cm,width=7cm]{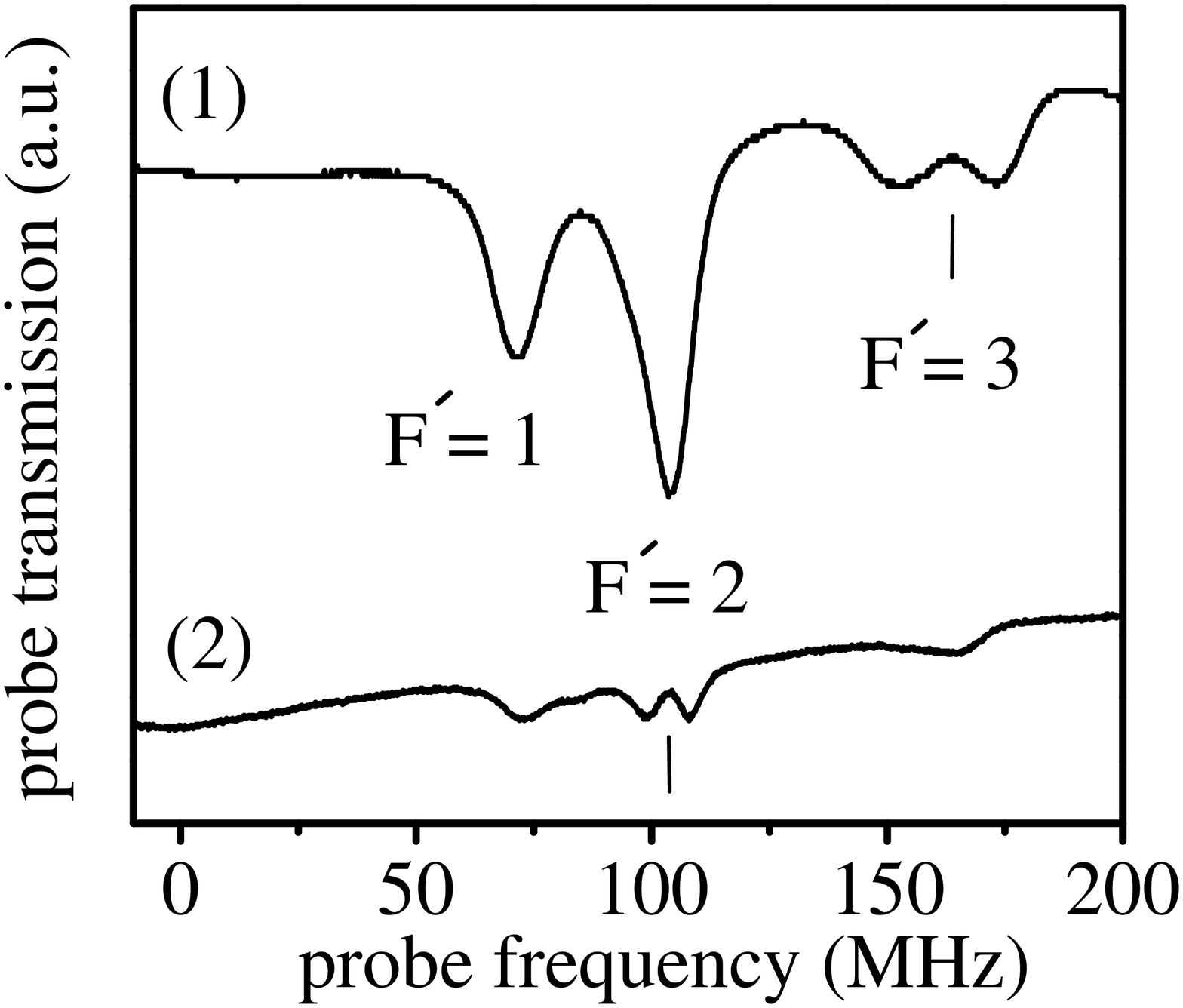}}
\hspace{0.2cm}
\subfigure[]{
\includegraphics[height=6.5cm,width=7cm]{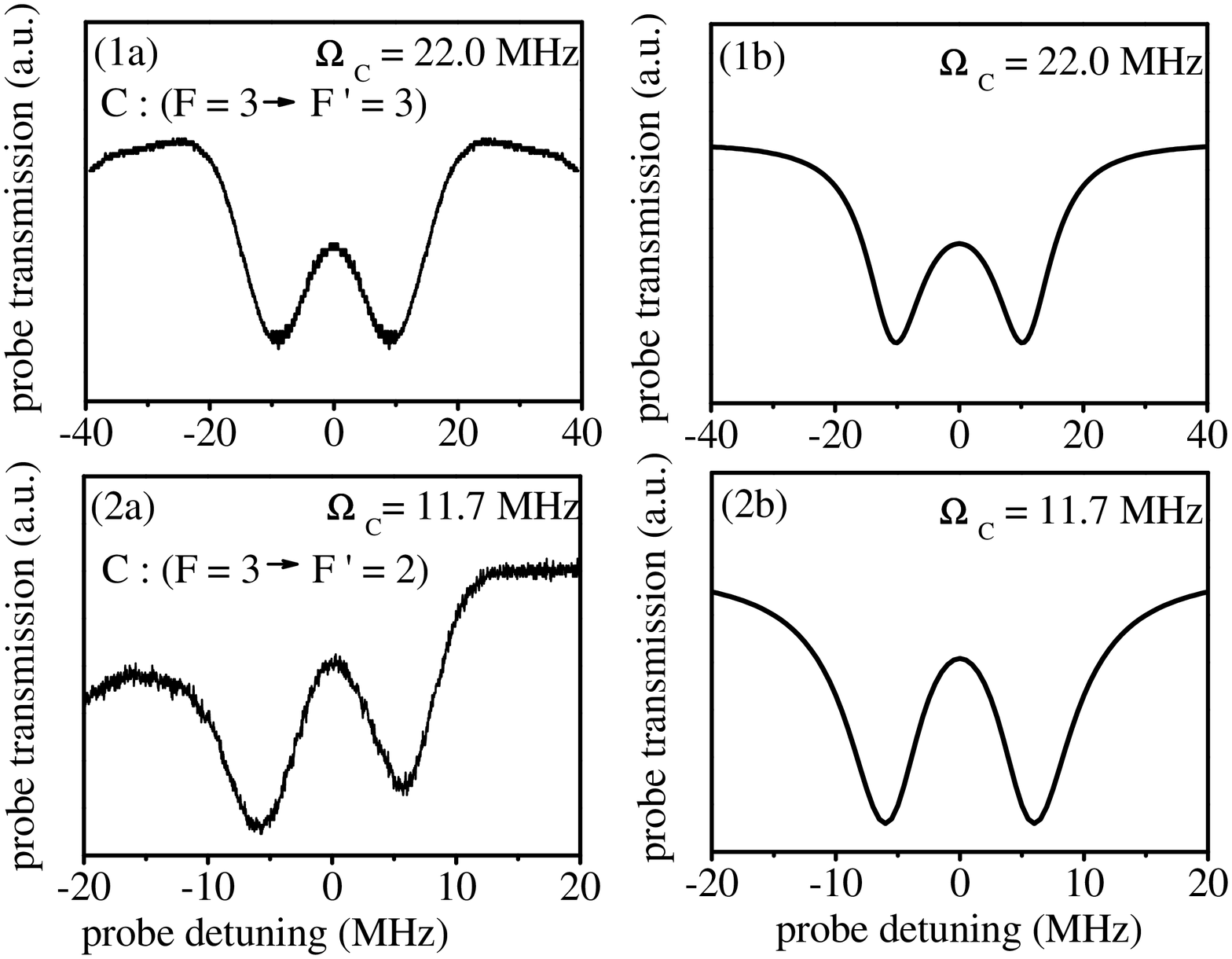}}
\end{center}
\caption{ Probe transmission spectra corresponding to the $D_{2}$ line of $^{85}$Rb for the same value of coupling power. (a) Curve (1): The coupling field with Rabi frequency $\Omega_{c}$ = 22.0 MHz, tuned to exact resonance with the 5 $^{2}S_{1/2}$ (F = 3) $\to$ 5$^{2}P_{3/2}$ (F'= 3) transition. Curve (2): The coupling field with Rabi frequency $\Omega_{c}$ = 11.7 MHz, tuned to exact resonance with the 5 $^{2}S_{1/2}$ (F = 3) $\to$ 5$^{2}P_{3/2}$ (F'= 2) transition. EIT signals are shown by vertical lines. (b) The curves in (1a) and (2a) are the close-ups of the experimental EIT signals mentioned above. The corresponding theoretically calculated curves are shown in (1b) and (2b), respectively.}
\end{figure}

\begin{figure}[htb]
\begin{center}
\subfigure[]{
\includegraphics[height=6.5cm,width=7cm]{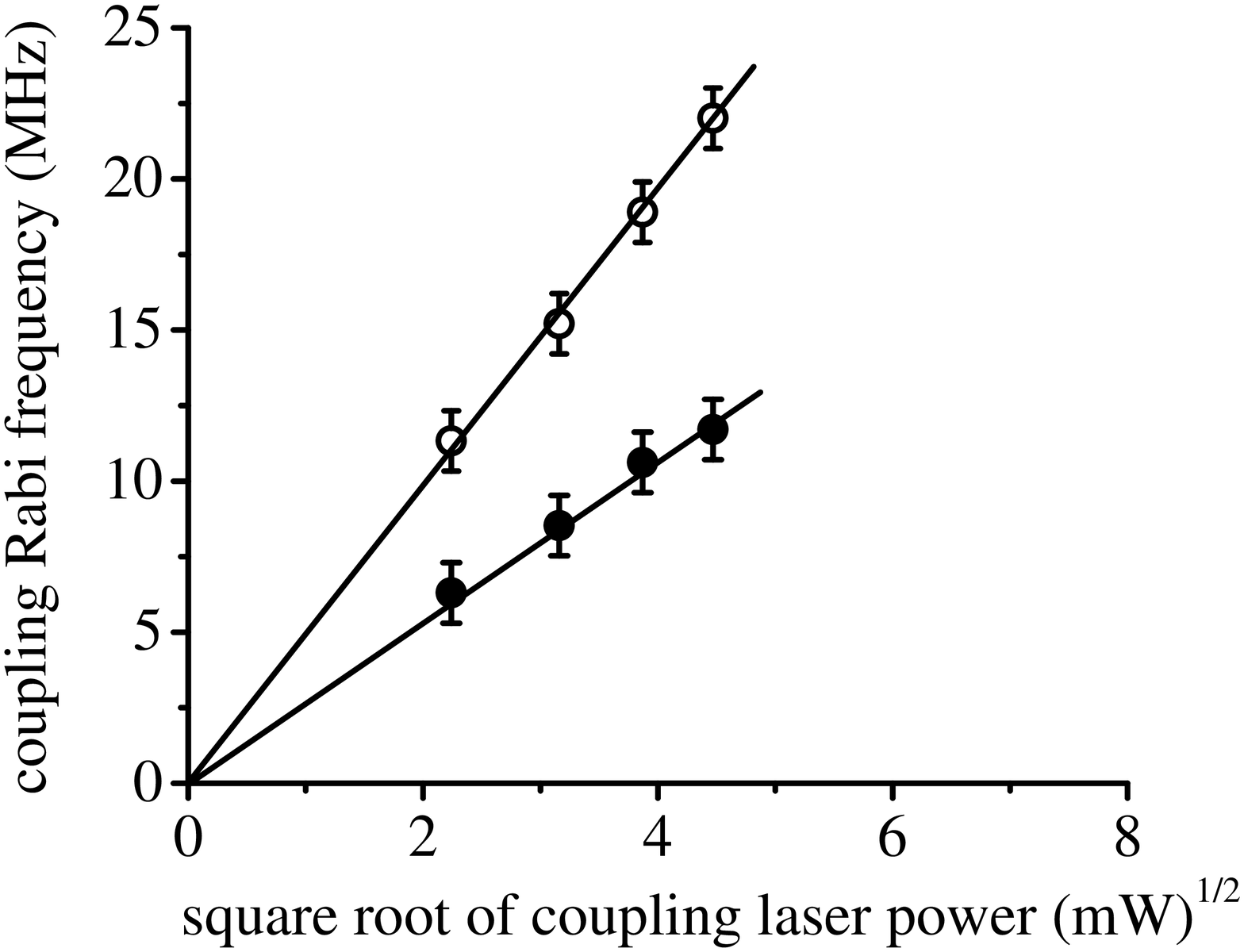}}
\hspace{0.2cm}
\subfigure[]{
\includegraphics[height=6.5cm,width=7cm]{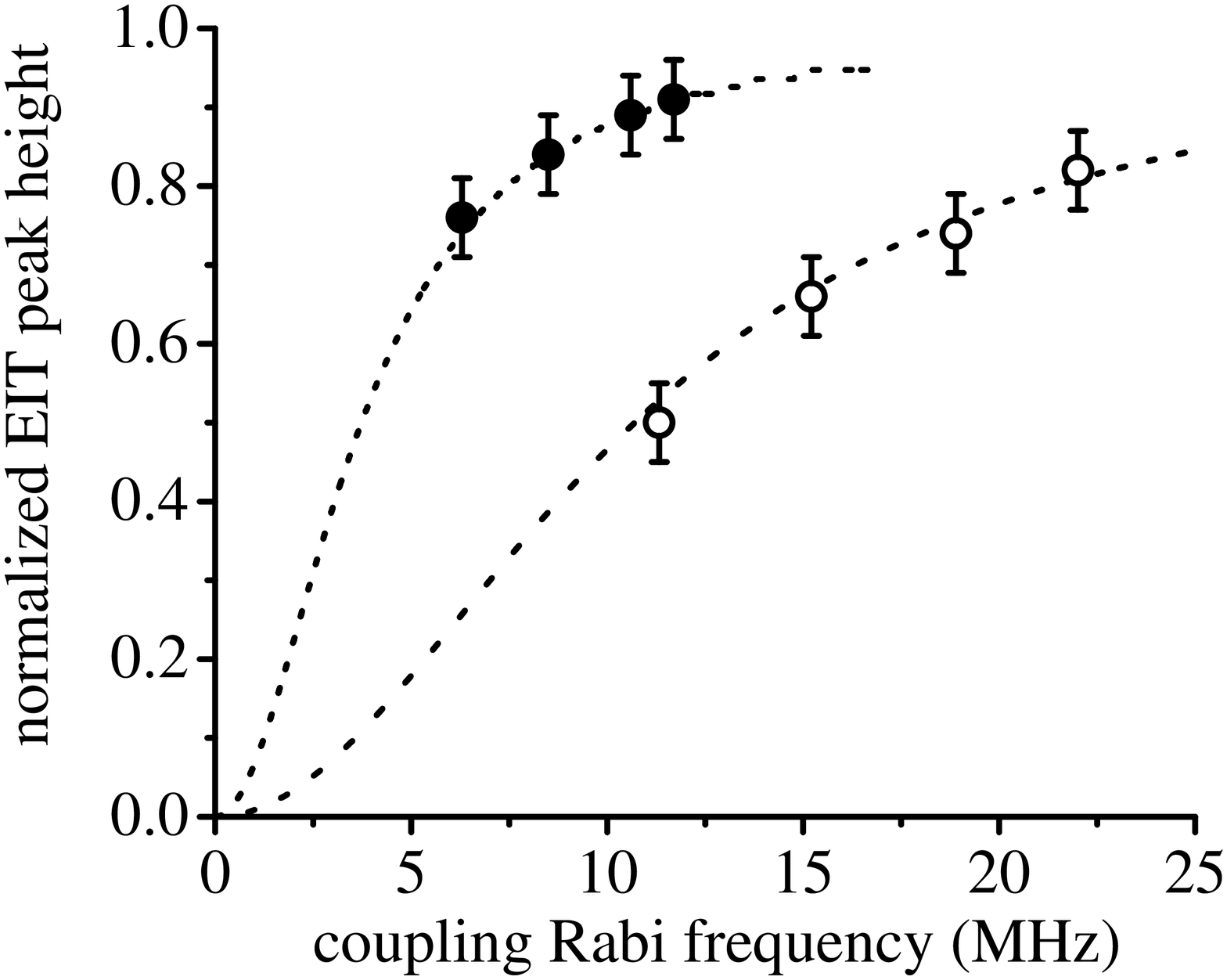}}
\end{center}
\caption{(a) Dependence of measured Rabi frequency on coupling laser power for EIT observed in cold $^{85}$Rb atoms trapped in F=2 level; Open circles are the experimental data for coupling to F'=3 level. Closed circles are the experimental data for coupling to F'=2 level. Solid lines represent theoretical calculations. (b) Normalized EIT peak height versus the coupling Rabi frequency $\Omega_{c}$ in the $\Lambda$-type system. Open circles are the experimental data for coupling to F'=3 level. Closed circles are the experimental data for coupling to F'=2 level. Dotted lines represent theoretical calculations.}
\end{figure}

\end{document}